\documentclass[aps,prl,twocolumn,showpacs,groupedaddress]{revtex4-1}

\usepackage[centertags]{amsmath}
\usepackage{graphicx}
\usepackage{amsfonts}
\usepackage{epstopdf}
\usepackage{ulem}
\usepackage{color}
\usepackage{array}
\usepackage{multirow}
\usepackage{ulem}
\usepackage{bbding}

\begin{document}
\title{Super rogue waves in simulations based on weakly nonlinear and fully nonlinear hydrodynamic equations}
\author{A.~Slunyaev$^{1,2,\ast}$, E.~Pelinovsky$^{1,2,3}$, A.~Sergeeva$^{1,2}$, A.~Chabchoub$^{4,5}$, N.~Hoffmann$^{4,5}$, M.~Onorato$^{6,7}$, and N.~Akhmediev$^8$}
\affiliation{$^1$ Institute of Applied Physics, N. Novgorod, Russia}
\email{Slunyaev@hydro.appl.sci-nnov.ru}
\affiliation{$^2$ Nizhny Novgorod State Technical University, N. Novgorod, Russia}
\affiliation{$^3$ Johannes Kepler University, Linz, Austria}
\affiliation{$^4$ Mechanics and Ocean Engineering, Hamburg University of Technology
Ei{\ss}endorfer Stra{\ss}e 42, 21073 Hamburg, Germany}
\affiliation{$^4$ Department of Mechanical Engineering, Imperial College London, London SW7 2AZ, United Kingdom}
\affiliation{$^5$ Mechanics and Ocean Engineering, Hamburg University of Technology, 21073 Hamburg, Germany}
\affiliation{$^6$ Dipartimento di Fisica, Universit\`a degli Studi di Torino, Torino 10125, Italy}
\affiliation{$^7$ Istituto Nazionale di Fisica Nucleare, INFN, Sezione di Torino, 10125 Torino, Italy}
\affiliation{$^8$ Optical Sciences Group, Research School of Physics and Engineering, The Australian National University, Canberra ACT 0200, Australia}

\begin{abstract}
The rogue wave solutions (rational multi-breathers) of the nonlinear Schr\"odinger equation (NLS) are tested in numerical simulations of weakly nonlinear and fully nonlinear hydrodynamic equations. Only the lowest order solutions from 1 to 5 are considered.
A higher accuracy of wave propagation in space is reached using the modified NLS equation (MNLS) also known as the Dysthe equation. This numerical modelling
allowed us to directly compare simulations with recent results of laboratory measurements in \cite{Chabchoub2012c}. In order to achieve even higher physical accuracy, we employed fully nonlinear simulations of potential Euler equations.
These simulations provided us with basic characteristics of long time evolution of rational solutions of the NLS equation in the case of near breaking conditions. The analytic NLS solutions are found to describe the actual wave dynamics of steep waves reasonably well.
\end{abstract}

\maketitle

\section{Introduction}
Rogue waves in the ocean is an intriguing geophysical problem, which has attracted much interest in the recent years (see reviews \cite{Dysthe2008,Kharif2009,Slunyaev2011}). Although quasi-linear random wave superposition in principle can generate extremely high waves, it is the nonlinear effect of the Benjamin-Feir instability of deep-water surface waves which is now believed to be the main reason why the rogue wave occurrence is so high in comparison with predictions of the basic Rayleigh theory
\cite{onorato01,ONO06}. The intermediate stage of the instability is associated with generation of phase-correlated structures – nonlinear wave groups, similar to that known within the framework of the integrable nonlinear Schr\"odinger (NLS) equation \cite{Zakharov1968,Zakharov1972,Osborne2010}. The NLS equation is a very basic first approximation for the sea wave dynamics in the limit of small wave steepness and long wave modulations. It is amazing that the weakly nonlinear analytic solutions governed by the NLS turned out to be well reproducible in laboratory conditions even for the steep (almost breaking) waves \cite{Chabchoub2012c,Chabchoub2011,Chabchoub2012a,Slunyaev2012}.

One class of the NLS solutions has particular interest, since they are simple enough to be derived analytically, and at the same time they exhibit outstanding capability of wave enhancement. The Peregrine breather is the first wave solution among this hierarchy, which has become recognized as the general prototype of rogue waves on the background of ordinary waves \cite{Peregrine1983,Shrira2010}. It describes the growth of infinitesimal perturbation of a plane wave, which culminates in the amplification that is three times the initial wave amplitude; after a fleeting stage of appearance, the rogue wave disappears and the plane wave is restored again (at least within the NLS equation framework). More general kinds of breather solutions were discovered in \cite{Kuznetsov1977,Akhmediev1985,Akhmediev1987}.

The hierarchy of higher-order rational breather solutions has also been found \cite{Akhmediev1985,Akhmediev2009a,Akhmediev2009b,Dubard2010,Gaillard2012}. They describe a nonlinear superposition of a group of rogue waves, which is localized both in time and space; they correspond to the case of degenerate eigenvalues of the associated scattering problem for the NLS. It was found \cite{Akhmediev2009b} that the maximum wave amplification described by the solutions is given by the formula $2N+1$, where $N$ is the order of the rational solution. The Peregrine breather corresponds to $N=1$. Observation of rational solutions of the NLS equation in a laboratory has been reported in a series of publications \cite{Chabchoub2011,Chabchoub2012a,Chabchoub2012b}.
 Such waves have been named {\it super rogue waves}. Up to $N=5$ solutions were successfully reproduced in a wave flume \cite{Chabchoub2012c}. Remarkably, the fifth-order solution exhibits the maximum wave amplification that is 11 times above the level of the background wave.

The experiments in \cite{Chabchoub2012c} demonstrated a good agreement between the analytic results given by the NLS solutions and the experimental observations. The agreement was achieved in both the dynamics of modulated waves and the observed shapes of the focused wave groups. The boundary condition for the wavemaker at $x=-d$ was specified in the form $A_N\left(-d, t\right)$, where $A_N$ is the analytic solution of the NLS equation, which corresponds to the $N$-order rational breather, and which attains the maximum amplification at the position $x=0$ in the moment of time $t=0$. Since the wave tank has limited length $L$, the following inequality $\left| d \right| < L$ must be satisfied to enable measuring of the wave profile at the location of the maximum amplitude (which corresponds to $x=0$ within the NLS framework). Indeed, the wave profiles measured at a distance $\approx d$ downstream (near $x=0$) were close to the analytic prototypes even when the focused wave was near the breaking point. This fact allows us to claim that the analytic rational breather solutions capture the nonlinear effects well and can be considered as a good first approximation.

At the same time, limitations of the experimental facility (limited length of the flume, gauge accuracy, capillary and dissipation effects) resulted in the use of rather short distances of the wave evolution ($d$ is small). The wave growth occurs in the scale of $\varepsilon^{-2}$ where $\varepsilon=ka$ is the wave steepness of the background wave. This is relatively long process in comparison with the propagation time along the tank. Since the distance $d$ is small, the perturbation generated by the wavemaker is not infinitesimal but has finite amplitude. The wave enhancement from one side of the tank to the focusing point is only a part of the total amplification $2N+1$ predicted by the theoretical solution on a long distance.

In the present study we analyse longer distances of wave propagation than those achievable in a water tank. This is done using numerical simulations rather than actual experiments. Namely, we solve partial differential equations that are more accurate in describing water waves than NLS.
Current results complement those presented in \cite{Chabchoub2012c}.
As in the experiment, only rational solutions of orders $N=1, 2, 3, 4, 5$ are considered. The techniques and analytic expressions were given earlier in \cite{Akhmediev2009b,Dubard2010,Gaillard2012}. The actual expressions are cumbersome and cannot be reproduced here. Two main questions posed in our previous work and addressed here are:

\begin{enumerate}

\item How longer distances/times influence the nonlinear wave focusing in realistic conditions of steep waves (larger $d$ and, correspondingly, focusing time $\tau$)? Do analytic solutions of the NLS equation provide adequate models for description of nonlinear wave focusing of intense near breaking waves?
    \textcolor{red}{Clearly, if the breaking does happen either for very high initial amplitudes or after large distances of propagation, the wave description requires different analysis which is beyond the theory and simulations in the assumption of smooth water surface}.

\item What is the profile of the amplified wave near maximum. Does the intensity of the background waves changes, when the focused waves start to break? The wave breaking onsets for breathers of different orders were estimated in the experiments \cite{Chabchoub2012c}. Namely, higher-order breathers $N=5$ with extremely small steepness of the background wave ($ka>0.01$) were found to be breaking due to the nonlinear focusing.
\end{enumerate}

We examine the observations made in laboratory experiments \cite{Chabchoub2012c} and compare them firstly with the results of improved model for nonlinear wave modulations presented in Section 2 and secondly with fully nonlinear simulations given in Section 4. These techniques allowed us to make a better modelling of breathers appearing on top of surface gravity waves. We paid special attention, in Section 3, to the wave enhancements reached by the nonlinear wave focusing in the weakly and strongly nonlinear approaches. Main outcomes of the study are summarised in the Conclusions.

\section{Wave evolution within the framework of the Dysthe equation}

The first-order approximation for deep-water surface gravity waves is the NLS equation \textcolor{red}{\cite{Zakharov1968,Osborne2010}.
It can be written in the form:}
\begin{equation}\label{nlse}
i\left(\frac{\displaystyle\partial A}{\displaystyle \partial
t}+C_{gr}\frac{\displaystyle\partial A}{\displaystyle \partial
x}\right)+\frac{\displaystyle \sqrt{gk_0}}{\displaystyle
8k_0^2}\frac{\displaystyle\partial^2A}{\displaystyle \partial
x^2}+\frac{\displaystyle \sqrt{gk_0}}{\displaystyle 2} k_0^2 \left|A\right|^2A=0.
\end{equation}
Parameters $k_0$ and $\omega_0$ are carrier wavenumber and cyclic frequency, which are linked
according to the deep-water dispersion relation $\omega_0^2= gk_0$.
The constant $g$ is the gravity acceleration, while the group velocity is given by $C_{gr}=\frac{\displaystyle\omega_0}{\displaystyle 2k_0}$.
The first approximation for the water surface elevation $\eta(x,t)$ is then given by $$\eta=\textnormal{ Re }[A(x,t)\exp(i\omega_0t- ik_0x)].$$

More convenient for direct comparison with laboratory experiments are evolution equations that describe wave propagation in space.
%Using linear relations involving the group velocity, the NLS (\ref{nlse}) can be transformed to the following form
\textcolor{red}{When the two first terms in (\ref{nlse}) which describe wave transport are assumed to be of the leading order, the NLS equation (\ref{nlse}) can be transformed to the following form}
\begin{equation} \label{nlse2}
i\left(\frac{\displaystyle\partial A}{\displaystyle \partial
x}+\frac{\displaystyle 1}{\displaystyle C_{gr}}\frac{\displaystyle\partial A}{\displaystyle \partial
t}\right)+\frac{\displaystyle 1}{\displaystyle
g}\frac{\displaystyle\partial^2A}{\displaystyle \partial
t^2}+\frac{\displaystyle\omega_0^6}{\displaystyle g^3}\left|A\right|^2A=0.
\end{equation}
\textcolor{red}{The details of the transformation from Eq.(\ref{nlse}) to (\ref{nlse2}) can be found for example in Chapter 12 of \cite{Osborne2010} or in \cite{Trulsen2006}. When keeping accuracy within the NLS approximation,}
 Eq.(\ref{nlse2}) is equally valid with (\ref{nlse}). The latter form is predominantly used in optics \textcolor{red}{while the form (\ref{nlse}) is more popular in the water wave community. However, for the sake of comparisons with the results obtained below, we will need the form (\ref{nlse2})}.

In this Section, we use the so called Dysthe model to simulate wave dynamics more realistically. This is
the modified nonlinear Schr\"odinger (NLS) equation (MNLS), which takes into account higher-order terms responsible for nonlinear dispersion, full dispersion of linear waves and the effect of induced long-scale current \cite{Dysthe1979,Trulsen1996}. Duality of the form of the classic NLS equation (\ref{nlse}), (\ref{nlse2}), allows us to write down either temporal and spatial versions of the MNLS. For convenience, we use here the version of the MNLS equation, which describes the wave evolution in space
\begin{align}\label{mnls}
&i\left(\frac{\displaystyle\partial A}{\displaystyle \partial
x}+\frac{\displaystyle 1}{\displaystyle C_{gr}}\frac{\displaystyle\partial A}{\displaystyle \partial
t}\right)+\frac{\displaystyle 1}{\displaystyle
g}\frac{\displaystyle\partial^2A}{\displaystyle \partial
t^2}+\frac{\displaystyle\omega_0^6}{\displaystyle g^3}\left|A\right|^2A\\ \nonumber
&-8i\frac{\displaystyle\omega_0^5}{\displaystyle g^3}\left|A\right|^2\frac{\displaystyle\partial A}{\partial t}-2i\frac{\displaystyle\omega_0^5}{\displaystyle g^3}A^2\frac{\displaystyle\partial A^*}{\displaystyle\partial t}-\frac{\displaystyle4\omega_0^4}{\displaystyle g^3}A\frac{\displaystyle\partial\varphi}{\displaystyle\partial t}\Big\vert_{z=0}=0,
\end{align}
where:
\begin{align*}
\frac{\displaystyle\partial\varphi}{\displaystyle\partial z}=-\frac{\displaystyle\omega_0}{\displaystyle 	2 C_{gr}}\frac{\displaystyle\partial}{\displaystyle\partial t}\left|A\right|^2, & \textnormal{ for } z=0;\\
\frac{1}{C_{gr}^2} \frac{\displaystyle\partial^2\varphi}{\displaystyle\partial t^2}+\frac{\displaystyle\partial^2\varphi}{\displaystyle\partial z^2}=0, & \textnormal{ for } z\leq\eta;\\
\frac{\displaystyle\partial\varphi}{\displaystyle\partial z}\longrightarrow 0, & \textnormal{ for } z \longrightarrow\infty.
\end{align*}
\textcolor{red}{The MNLS equation is obtained in the order higher than validity of the NLS equation. It includes the terms of order $O(\epsilon^4)$. As another improvement, it takes into account the effect of induced mean flow, described by the function $\varphi(x, z, t)$. Here, an additional variable $z$ is the upward directed vertical axis. The real-valued function $\varphi$ satisfies the Laplace equation in the water column with the boundary conditions at $z = 0$ and $z \rightarrow -\infty$. The details of this approach can be found in \cite{Dysthe1979}, \cite{Slunyaev2005} and \cite{Trulsen2006}.}

Just like in the laboratory runs, we use higher-order rational analytic solutions of the NLS equation $A_N(x= -d, t)$, $N=1, 2, 3, 4, 5$, to specify the boundary condition for the MNLS simulations starting  at $x= -d$. Note that two different forms of the NLS equation (1) and (2) allows one the same solution to be written twofold. In order to start simulations of the spatial version of the MNLS equation (\ref{mnls}), we
use the corresponding solutions of the spatial version of the NLS equation (\ref{nlse2}). The programming code of the Dysthe model used here has been verified earlier in the laboratory studies of intense water wave packets \cite{Shemer2010}.

Similar to the classic NLS equation, the Dysthe model (3) governs the evolution of the complex wave amplitude $A(x,t)$. Therefore it is a straightforward task to compare them with solutions of the NLS equation. As we used pseudospectral method to simulate the solutions of Dysthe equation, the time domain must have  periodic boundary conditions. Although the breathers are defined on the infinite domain, they are localised and we used time intervals much wider than the characteristic localisation interval.

Time series should be used to start the simulations with evolution in space, which is governed by equations (\ref{mnls}). These time series can be easily calculated using analytic expressions and are not given here. To give an idea, their wave profiles are similar to the space series shown in Fig.4 except that independent variable is $t$.
For consistency with experimental results, the wave amplitudes and periods as well as the associated wave lengths
are chosen in accordance with those used in laboratory experiments \cite{Chabchoub2012c}.
All essential experimental parameters are given in Table \ref{tab1}.

\begin{table}[ht]
\centering\begin{tabular}{ | >{\centering}p{2cm} | >{\centering}p{2cm} | >{\centering}p{2cm} | >{\centering\arraybackslash}p{2cm} |}
\hline
Order of rational solution, $N$ & Steepness of the unperturbed wave, $k_0A_0$ & Amplitude of the unperturbed wave $A_0$ (m) & Carrier wavelength $\lambda=\frac{2\pi}{k_0}$ (m) \\ \hline
    1 & 0.117 & 0.01 & 0.54 \\ \hline
    2 & 0.05 & 0.003 & 0.38 \\ \hline
    3 & 0.04 & 0.002 & 0.31 \\ \hline
    4 & 0.03 & 0.003 & 0.67 \\ \hline
    5 & 0.02 & 0.002 & 0.67 \\ \hline
\end{tabular}
\caption{Parameters for simulations of the MNLS equation.\label{tab1}}
\end{table}

Having more freedom in conducting numerical experiments we used significantly longer distances of propagation.
Instead of the limiting length of $d=9$ m in real experiment \cite{Chabchoub2012c} we used of up to 100 m in simulations.
As can be seen from Table 1, the distance $d=100$~m corresponds to the number of 150 -- 320 carrier wave lengths. The results of these simulations are shown in Fig.1. Clearly, the wave amplification is the most fascinating feature of the breather evolution. The evolution of the maximum of the complex wave amplitude, i.e., max~$\left|A\right|$, as a function of $x$ is shown in Fig.1. This is done for five different orders of the rational breather, from $N=1$ to $N=5$.  Each simulation is started at the distance $d$ to the left from the $x=0$ point with $d=9,~20,~30,~50$  and $100$ m. The coding for each curve is shown in the left hand side insets. For comparison, the maximum amplitude of the exact breather solution is shown by the thin black curve.

% Figure 1
\begin{figure}[h]
\centering
\includegraphics[width=8cm]{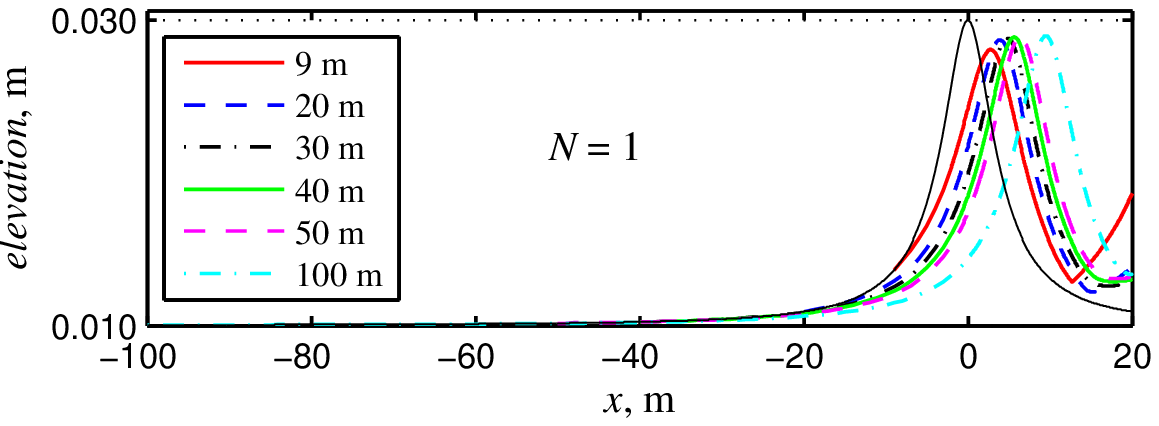}
\includegraphics[width=8cm]{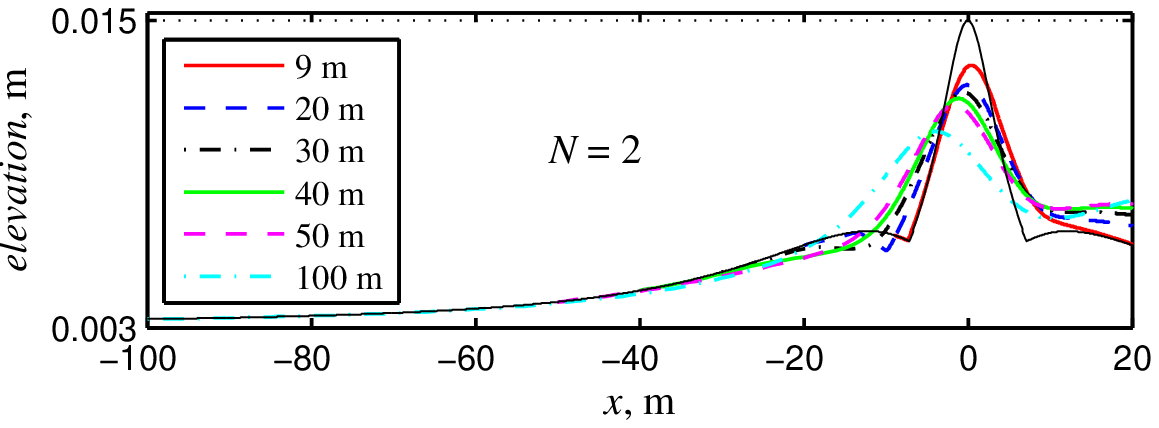}
\includegraphics[width=8cm]{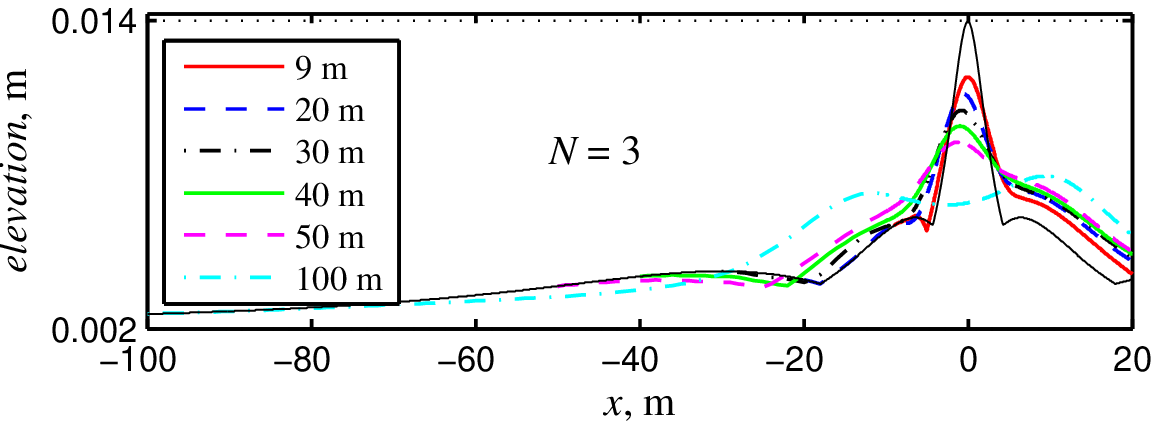}
\includegraphics[width=8cm]{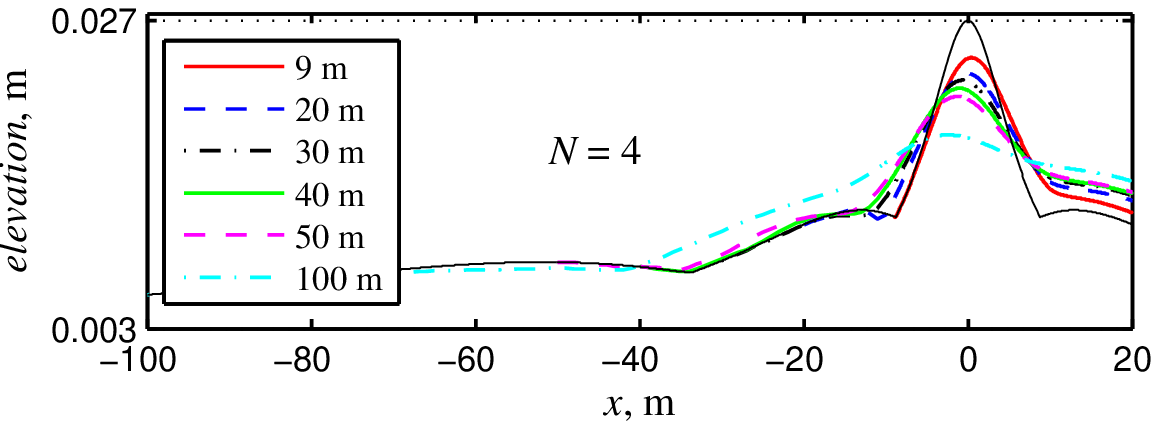}
\includegraphics[width=8cm]{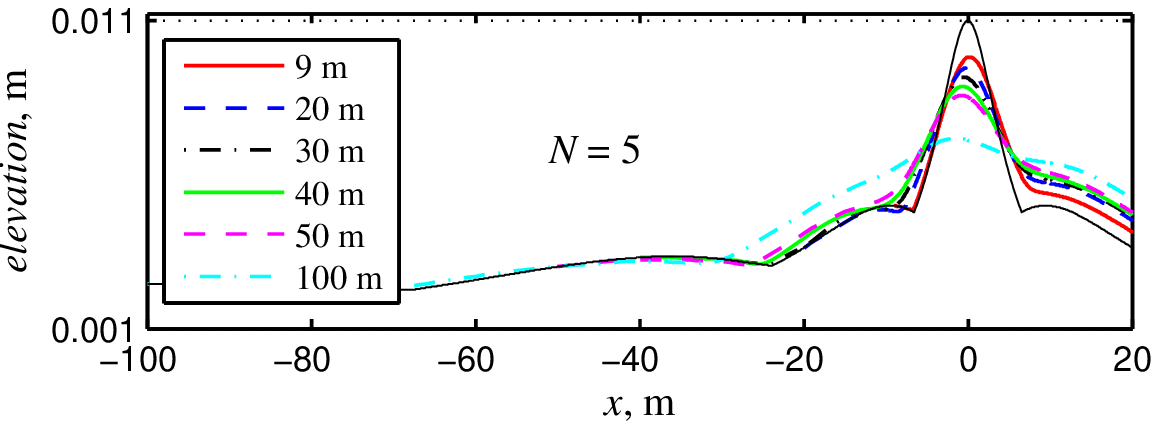}
\caption{(Color online) Evolution of the maximum amplitude $\left|A\right|$ in numerical simulations of the Dysthe model. The insets show the distance $d$ of propagation for each case before reaching the focusing point at $x=0$. A thin black line in each panel corresponds to the analytic solution for the rational breathers of the NLS.
The lower limit along the vertical axis shows the amplitude of the unperturbed background wave, $A_0$. The experimental parameters that correspond to each case are given in Table 1.}\label{fig1}
\end{figure}

A close examination of Fig.1 shows that despite a somewhat lower amplification observed in numerical simulations, the curve shapes are quite similar to those given by the NLS solution (thin solid line in each panel). The difference becomes noticeable when $N\geq 2$ and at $d$ values higher than $50$ m. This difference between the analytic NLS solution and the numerical solution of the Dysthe equation becomes qualitative in the case of the largest distance of $d=100$ m. A substantial deviation in Fig.1 can be observed for  the orders 3,~4 and 5.

Parasitic wave modulations grow along with the wave evolution. They are mainly located at the ends of the background wave train which is unavoidable feature of the technique and irrelevant to the breather itself. Examples of these parasitic modulations can be seen in Fig.5 which is done for fully nonlinear simulations.
The effect is very similar in the case of weakly nonlinear simulations of the Dysthe model.
	
In order to eliminate this effect,  the wave shifts close to the central peak should be carefully analysed when computing the maxima of $\left| A \right| $. Other unstable modes may also grow in the central area resulting in developing additional maxima in Figs.1. These maxima may appear at distances different from the expected location $x=0$ of the wave focusing.

Another quite visible feature of Fig.1 is that the case $N=1$ is qualitatively different from $N=2,...,5$.
There is a clear trend that at moderate values of $d$ the amplification grows with $d$ in the case $N=1$.
Then the starting conditions which correspond to larger $d$ result in a stronger wave amplification.
When $N=1$, even the longest distance $d=100$~m results in strong wave amplification.
This is not the case when $N \geq 2$.
When $N=1$, the maximum wave focusing actually occurs later than predicted by the NLS theory. Moreover, the delay increases with the distance $d$. On the other hand, there is no particular order in location of the focusing position in the cases of $N \geq 2$ although generally the maximum is closer to $x=0$.

Thus, the NLS equation provides good qualitative evolution of the point of maximum in the case of $N=1$ but predicts its appearance earlier than it happens in the Dysthe model. The amplification values of $\approx3$ for the Peregrine breather predicted in each model are in very good agreement. We should not forget that the starting boundary condition with small deviation from the constant amplitude background wave in simulations is the exact solution of the NLS equation. It may happen that corrections of this function may improve the agreement between the two models.

\section{Wave amplification}

\textcolor{red}{ The results presented in Fig.1 may seem to be self-explanatory. However, there is an essential point that has to be taken into account when analysing them.}
\textcolor{red}{Namely, the amplitude $A$ and the surface elevation $\eta$ are different functions.
%Thus, the maxima of $A$ in Fig.1 do not necessarily coincide with the maxima of the $\eta$-function.
Thus, the maxima of $A$ in Fig.1 and the maxima of the $\eta$-function may attain significantly different values.
According to the Dysthe theory, the surface elevation is computed taking into account bound waves up to the third order. Namely,
\begin{eqnarray}\nonumber
\eta(x,t)=&-&\frac{k_0}{\omega_0^2} \frac{\partial}{\partial t} \varphi \Big{|}_{z=0}+ \Re \left(A \exp(i\omega_0 t - ik_0 x) \right)\\
\nonumber
&+&\frac{k_0}{2} \Re \left(A^2 \exp(i2\omega_0 t - i2k_0 x) \right) \\
\nonumber
&+&\frac{k_0}{\omega_0} \Im \left(A \frac{\partial A}{\partial t} \exp(i2\omega_0 t - i2k_0 x) \right) \\
&+&\frac{3k_0^2}{8} \Re \left(A^3 \exp(i3\omega_0 t - i3k_0 x) \right)
\label{newf}
\end{eqnarray}
This approach is standard for the Dysthe theory, the details can be found in \cite{Trulsen1996,Slunyaev2005}.
As a result, the curves in Fig.1 need quantitative analysis in relation to the maximal amplitudes achieved in propagation. For this aim, we use the standard definition of the wave amplification over the background or over the initial wave amplitude.}

In particular, one of the parameters calculated in our study is the wave amplification $\chi$ defined as the ratio of the maximal amplitude $A_{\max}$ to the amplitude of the background wave $A_0$, i.e. $\chi=A_{max}/A_0$. \textcolor{red}{Thus, this is the maximal amplification with respect to the infinitively far state. Since in numerical simulations we cannot operate with the waves at infinity, we introduced the second parameter, $\rho$ which considers the wave amplification with respect to the initial / boundary condition.  In order to analyse the dynamics of these parameters in more detail, we plotted them} in Fig.2. Namely, Fig.2 displays the value of amplification reached in the simulations of the Dysthe model. This figure also contains the results of the strongly nonlinear simulations described below as well as experimental data.

% Figure 2
\begin{figure}[ht]
\centering
\includegraphics[width=7.1cm]{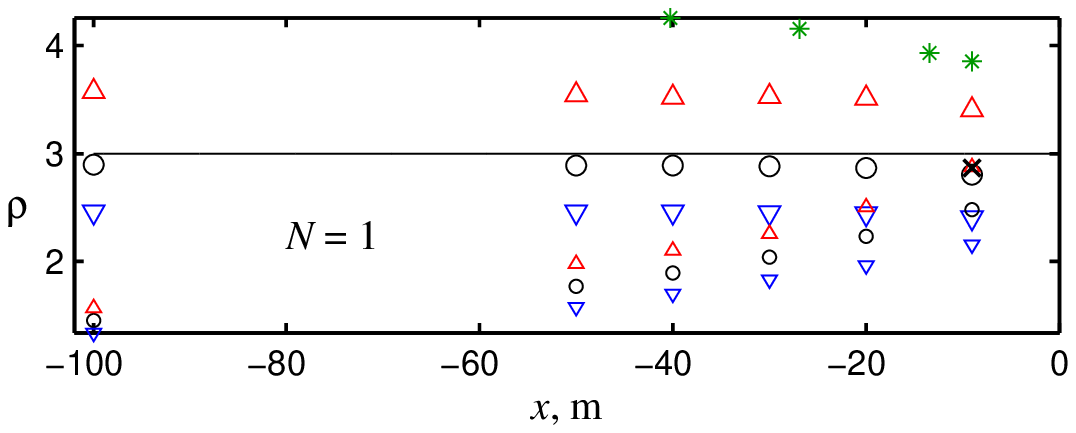} \vspace{-2mm}
\includegraphics[width=7.1cm]{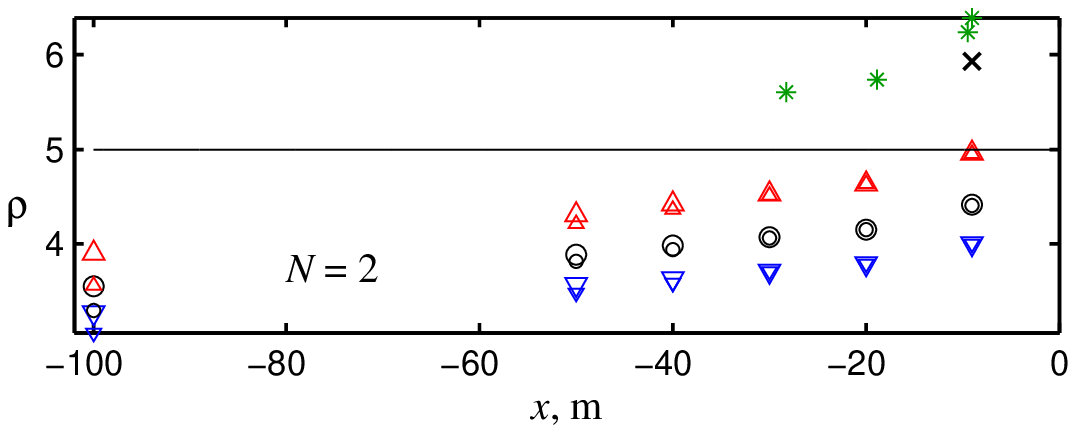} \vspace{-2mm}
\includegraphics[width=7.1cm]{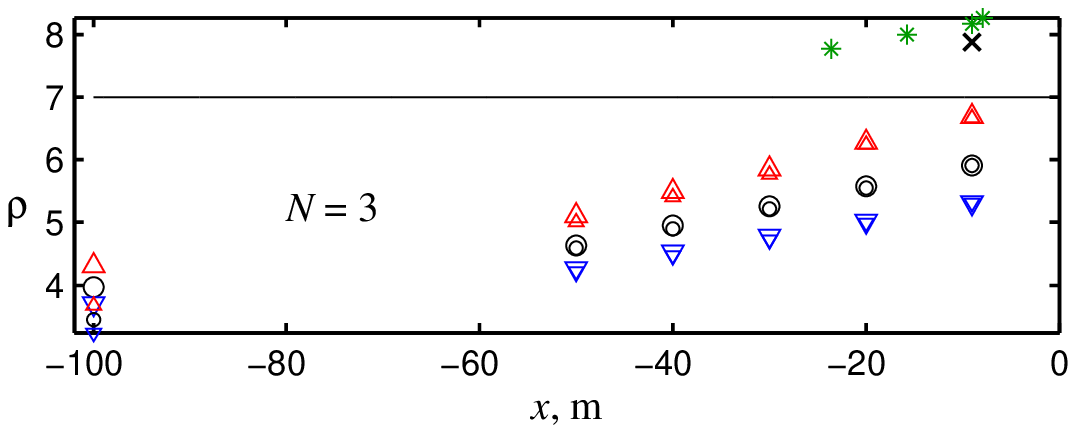} \vspace{-2mm}
\includegraphics[width=7.1cm]{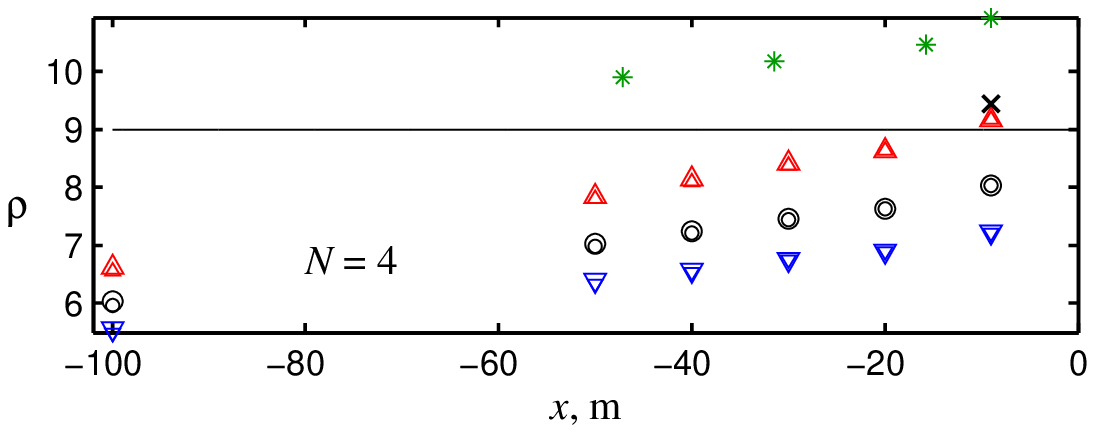} \vspace{-2mm}
\includegraphics[width=7.1cm]{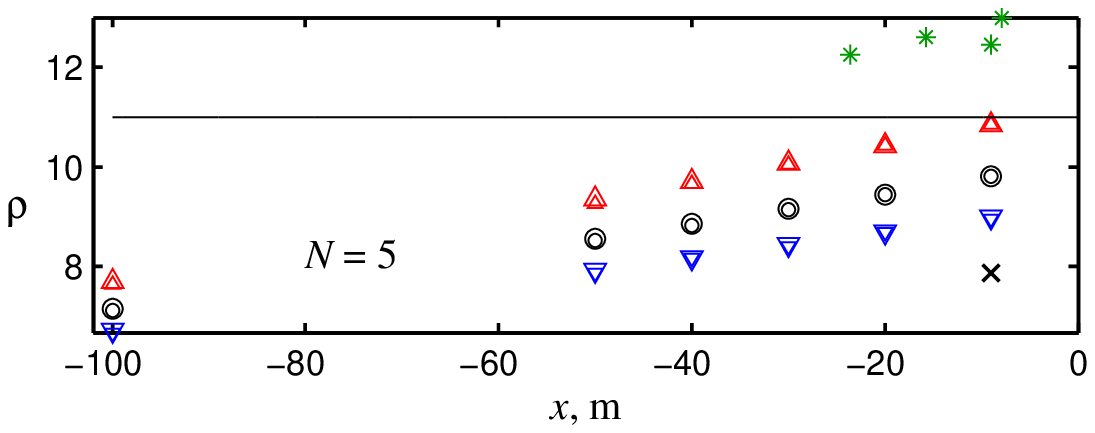} \vspace{-2mm}
\caption{(Color online) The relative values of wave maxima vs the distance $d$.
Notations are as follows. Simulations of the Dysthe model:
Wave amplification in terms of $\rho_1$ (circles \textcolor{black}{$\circ$}), $\rho_2$ (upward directed triangles \textcolor{red}{$\bigtriangleup$}) and $\rho_3$ (downward directed triangles \textcolor{blue}{$\bigtriangledown$}).
The larger size symbols among circles and triangles are the maxima attained within the simulated distance $\left(-d < x < 20 \textnormal{ m}\right)$. The smaller symbols of the same type denote the maxima in close proximity to the point $x=0$, typically within one wavelength.  Green asterisks (\textcolor{green}{$\ast$}) show the results of fully nonlinear simulations for near breaking conditions given in Section 4 and in the Table \ref{tab2} with the steepness values shown in bold. Crosses (\textcolor{black}{$\bf{\times}$}) correspond to the results of the experiment \cite{Chabchoub2012c}.
The horizontal line in each panel at the level of $2N+1$ is the amplification provided by the analytic NLS solution.
}\label{fig2}
\end{figure}

\textcolor{red}{Generally, the curves in Fig.1 may have maxima at locations different from $x = 0$. Moreover, the maxima of surface elevation $\eta$ are different from the maxima of $A$.
Correspondingly, we give two sets of the same symbols in the plots in Fig.2 -- the larger and the smaller ones. The larger symbols show the maximum amplification reached within the simulation domain $-d < x < +20$ m, while the smaller symbols represent only the amplification reached in the vicinity of the point $x=0$ closer than the wave length.}

The circles in these plots show the ratio $\rho_1$ of absolute maximum to the unperturbed envelope amplitude, i.e.
$$\rho_1=\max\left|A\right|/ A_0.$$
The upward (red) and downward (blue) directed triangles show the wave enhancement in terms of surface elevation and depression, i.e.,
$$\rho_2=\max(\eta)/ A_0$$
and
$$\rho_3=\min(|\eta|)/ A_0,$$
respectively. The surface elevation $\eta(x,t$) is computed taking into account bound waves up to the third order (\ref{newf}).
Horizontal lines in Fig.2 show the theoretical maxima $2N+1$ predicted by the NLS solutions. Generally, the complex wave amplitudes $\left|A\right|$ (circles) do not achieve the theoretical limit $2N+1$. However, quite remarkably,  the contribution of bound wave components results in a greater amplification of wave crests $\max(\eta)$. These values can even exceed the theoretical limit $2N+1$ (upward directed triangles).

% Figure 3
\begin{figure}[ht]
\centering
\includegraphics[width=7.9cm]{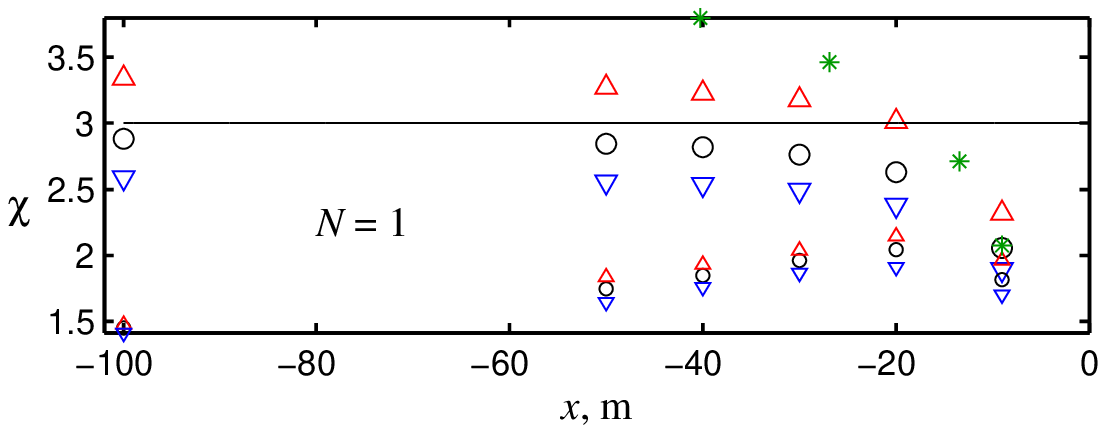}
\includegraphics[width=7.9cm]{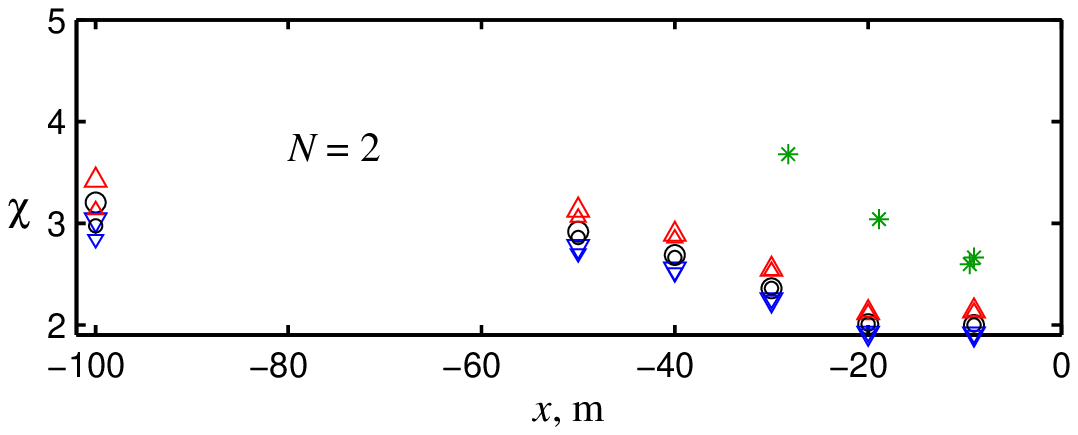}
\includegraphics[width=7.9cm]{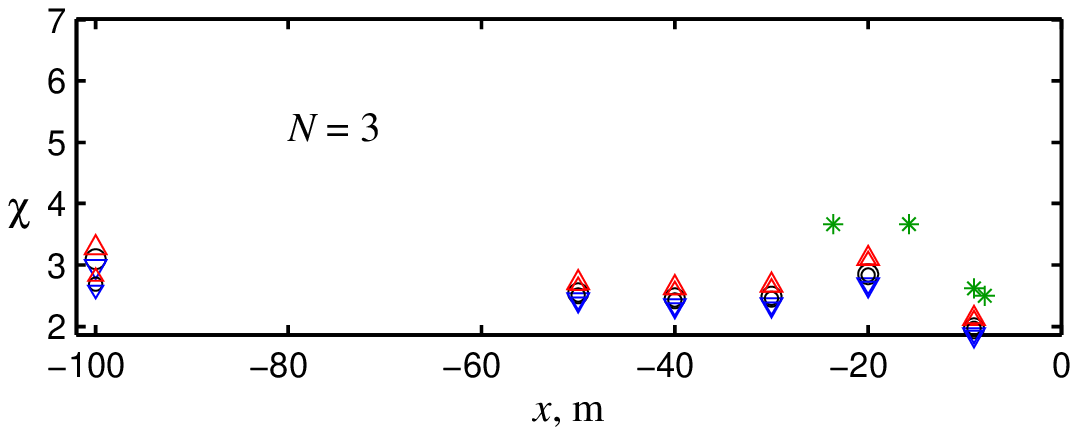}
\includegraphics[width=7.9cm]{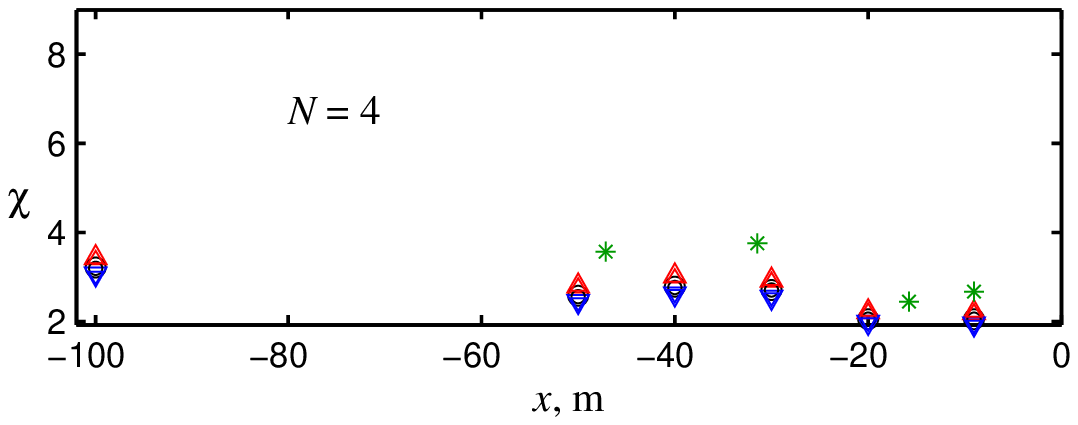}
\includegraphics[width=7.9cm]{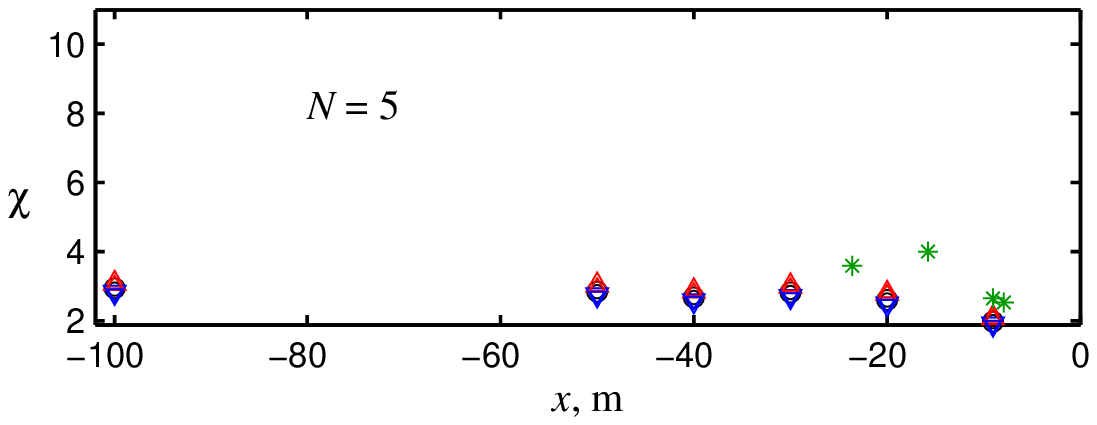}
\caption{(Color online) Same as Fig.2, but modified amplifications
$\chi_1$ (circles \textcolor{black}{$\circ$}), $\chi_2$ (upward directed triangles \textcolor{red}{$\bigtriangleup$}) and $\chi_3$ (downward directed triangles \textcolor{blue}{$\bigtriangledown$}) are shown along the vertical axis.
The upper limit of the vertical axis in the cases $N = 2, 3, 4, 5$
corresponds to the amplification provided by the corresponding analytic NLS solution, $2N+1$. Green asterisks (\textcolor{green}{$\ast$}) show the results of fully nonlinear simulations for near breaking conditions.}\label{fig3}
\end{figure}

When $N=1$, the focusing is delayed. Then the maximum which is measured at $x=0$ may be much smaller than after this point. Consequently, the smaller symbols show lesser amplification than the larger ones in the case $N=1$. On the other hand, when $N \geq 2$, the maximum wave amplitudes in the entire domain $-d < x < + 20$ m are similar to the ones near the point $x=0$.

As can be seen from Fig.2, for $N=1$ larger values of $d$ result in larger amplifications. Vice versa, when $N \geq 2$, larger amplifications are observed at smaller $d$. The results obtained in laboratory measurements \cite{Chabchoub2012c} are shown by crosses in Fig.2. In the experiment, $d=9$~m and this point can be compared with small upward directed triangles at the same distance $d$. The numerical and experimental points coincide in the case $N=1$ while the experimental points are slightly higher when $N=2, 3, 4$. When $N=5$, the experimental point is lower than the one obtained in numerical simulations. This can be attributed to significantly smaller steepness, $k_0 A_0 = 0.01$ in the laboratory experiment for $N=5$. We also note that the vertical scale in each panel does not start from zero and the differences are smaller than it may seem at first glance.

The results presented in Fig.2 clearly demonstrate that the maximum surface elevation at $x=0$ obtained for the Dysthe model with $d=9$ m is very close to the value which is $\left(2N+1\right)$ times the amplitude of the background wave. These are given by the  small upward directed triangles at $d=9$ m. Close data were indeed observed in the experiments \cite{Chabchoub2012c}. At the same time, the Dysthe model predicts the maximum wave amplification in the case of $N=1$ to occur at a larger distance than the NLS theory provides. This feature still has to be confirmed in laboratory tests.

The lower limit of the vertical axis in Fig.1 is the background wave amplitude $A_0= A_N \left(x\rightarrow \pm \infty \right)$.
 When $N$ is large, i.e. $N=3,4,5$, the wave train is already amplified with respect to the background wave amplitude $A_0$ at every $x$ before the focusing point. The amplification is noticeable even at the largest distance $d=100$ m considered in simulations, see Fig.1. If we use the real meaning of 'amplification', i.e. its value relative to the starting wave amplitude, the actual numbers would be smaller.
	
To take this correction into account, we calculated the amplifications with respect to the wave intensity at the starting point. Namely, we calculated the three following relative quantities
$$\chi_1=\frac{\max\left|A\right|}{\max \left|A_N \left(x=-d,t\right)\right|},$$
$$\chi_2=
 \frac{\max\left( \eta \right)} {\max \left( \eta_N \left(x=-d, t\right)\right)},$$
and $$\chi_3=\frac{\min\left(\eta\right) }{ \min\left(\eta_N\left(x=-d,t\right)\right)}$$
 instead of the those used in Fig.2. The new characteristic parameters are presented in Fig.3. Notations here are similar to those used in Fig.2, i.e. the circles, upward and downward directed triangles have the same meaning.
	
Clearly, if the initial amplitudes would coincide with the background one, Figures 2 and 3 would be identical.
The data in Fig.3 differ significantly from those in Fig.2 for short distances $d$ and most visibly for large values of $N$. For the shortest focusing distance $d=9$ m the modified wave amplification $\chi_i$ is around 2. The maximum wave amplification observed in all cases shown in Fig.3 is a little bit larger than three. Even though the NLS solution with $N \geq 2$ describes the dynamics better for shorter distances $d$, the overall wave enhancement is higher when $d$ is large. It is remarkable, that the first order solution exhibits systematically better agreement with the NLS exact solution when the distance $d$ increases. The higher-order cases, $N \geq 2$ do not have a clear trend.

% Section
\section{Rogue waves simulated in fully nonlinear Euler equations}

Simulations according to the Dysthe model provide the most convenient data for comparison with the laboratory experiments. However, the wave parameters given in Table 1 are very close to the breaking onset. The laboratory experiments \cite{Chabchoub2012c} show that the steeper waves are indeed breaking. Therefore, using the approximate Dysthe model still can be inadequate.

 In this Section, we give the results of the numerical simulations of the Euler equations. We assume potential fluid motions. Most of the numerical runs are performed using the Higher-Order Spectral Method (HOSM). The details of the numerical approach can be found in \cite{West1987}. These simulations resolve up to 7-wave interactions, i.e. we use $M= 6$. Thus, practically, our numerical experiments are fully nonlinear. Selected runs of near-breaking focused waves were duplicated with the solver of the Euler equations in conformal variables \cite{Chalikov,Euler}. The latter does not make any assumptions on wave steepness, and thus is a fully nonlinear approach. Comparing the two we can validate the results of the HOSM simulations.
 %We give here only the results of HOSM runs which did not show differences from the results provided by the direct simulations of Euler equations.
 Consequently, we do not distinguish the results obtained using the two numerical approaches.

The numerical codes have been developed previously in \cite{Slunyaev2009}. We are mainly interested in the strongly nonlinear case when the focused wave is close to breaking. Near-breaking waves have the greatest enhancement rate \cite{Slunyaev2012b}. The unstable mode which is the closest analog of the Peregrine breather with $N=1$ has the amplification which actually exceeds the value of four. Remarkably, this is higher than the original amplification of $3$.

The steepness  ($k_0A_0$) of the near-breaking NLS solution was estimated on the basis of laboratory measurements in \cite{Chabchoub2012c}. This data are shown in the last column of Table 2. Based on the results of the previous Sections, one may expect that this threshold should be universal for the case $N=1$. Indeed, the amplification does not depend very much on $d$. Moreover, it converges to a single value when $d$ increases.

For small distances $d$ and $N \geq2$, there is no universal steepness limit for wave breaking. This happens
because the amplification factor and the wave steepness depend strongly on $d$ as can be seen from Figs.2 and 3. For large $d$, the breaking criterion should be close to the instability limit of the $N=1$ case.

The Euler equations govern the evolution in time. The approach is qualitatively different from solving the NLS (\ref{nlse2}) or Dysthe (\ref{mnls}) equations. The initial wave profile along $x$ must be given. Posing the initial value problem results in the fully nonlinear solution.
We used several sets of initial conditions. \textcolor{red}{We followed the conditions of the  laboratory experiments \cite{Chabchoub2012c} taking
$A_N(x,t=-L/C_{gr})$ where $A_N$ is the analytic solution of (\ref{nlse}), $L=9$ m is the effective length of the tank and $C_{gr}$ is the carrier group velocity.
The surface elevation and the velocity potential at the initial instant are found taking into account bound waves of three asymptotic orders and the induced mean flow, similar to (\ref{newf}), see details in \cite{Slunyaev2009}.}
The focusing distance in terms of the wavelength, $\lambda_0$ and the wave period, $T_0$, is given in the second column of Table \ref{tab2}. The wavelengths $\lambda$ are given in Table 1.

As we have to deal with periodic boundary conditions, the total numerical grid has been chosen much larger than the space interval we are interested in. This can be seen in Fig.4. Various wave steepnesses $k_0A_0$ have been considered for each wavelength. The intention was to find the steepness value when the breaking occurs. The latter manifests itself in abrupt instability of numerical simulations.

The values of parameters when the wave evolution is either smooth or breaks down are given in the third column of Table \ref{tab2}. The cases of survived waves with smooth evolution are highlighted in bold. The cases with breaking are shown in normal fonts. Thus, the breaking threshold parameters are in between these values.
The breaking thresholds found in laboratory experiments \cite{Chabchoub2012c} are given in the last column of Table \ref{tab2}. Comparison of the last two columns in Table \ref{tab2} shows that the breaking limit in the case $N=1$ is much smaller in numerical simulations than in the corresponding laboratory measurements. One of the possible explanations is that  in the experiments waves were not monitored at distances longer than $d$ i.e. at $x>0$.

As can be seen from Figs.1-3 of the previous Section, the strongest wave amplification in the case $N=1$
 occurs at $x>0$. This observation agrees with the results of \cite{Slunyaev2012b}). Our fully numerical simulations were stopped either when the wave breaking had occurred or when the wave maximum had been reached.
Slightly lower estimates of the steepness threshold for wavebreaking may occur for values $N=3, 4$. However, in general, the fully nonlinear simulations capture the wave breaking onset rather well.
\begin{table}
\centering\begin{tabular}{ | >{\centering}p{1.0cm} | >{\centering}p{2.7cm} | >{\centering}p{1cm} | >{\centering}p{1.cm} | >{\centering\arraybackslash}p{2cm} |}
\hline
Order of ra-\\tional so-\\lution $N$ & Focusing \\distance $d$ / time $\tau$ & \multicolumn{2}{>{\centering}p{2cm} |}{Breaking onset range in numerical simulations $k_0A_0$}  & Breaking onset in laboratory experiments \cite{Chabchoub2012c} \\
    \cline{3-4}
       &      &  Fo-\\cused wave sur-\\vives  & Wave breaking occurs & \\ \hline
    1 & $16.76\lambda\approx33.52T_0$ & \bf{0.090} & 0.095 & 0.12\\ \hline
    2 & $23.87\lambda\approx47.74T_0$ & \bf{0.060} & 0.065 & 0.06\\ \hline
    3 & $28.65\lambda\approx53.30T_0$ & \bf{0.040} & 0.045 & 0.05\\ \hline
    4 & $14.32\lambda\approx28.64T_0$ & \bf{0.030} & 0.035 & 0.04\\ \hline
    5 & $28.65\lambda\approx53.30T_0$ & \bf{0.025} & 0.030 & 0.02\\ \hline
\end{tabular}
\caption{\label{tab2}Parameters of the fully nonlinear simulations.}
\end{table}

The second set of initial conditions corresponds to longer focusing times, when the initial conditions are chosen in the form $A_N\left(x,t=-\tau \right)$, where $\tau = 50T_0, 100T_0$ and $150T_0$. The carrier wavenumber in all cases is set to $k_0 = 20$~rad/m, and the steepness is given in bold fonts in Table \ref{tab2}.

The amplifications observed in the fully nonlinear numerical simulations are shown in Figs.2 and 3 with asterisks. Finding the amplification value needs an appropriate recalculation of the distance according to the difference in the wavelengths of the carrier. In all cases the amplification values exceed the results of weakly nonlinear simulations of the Dysthe model but clearly follow the trends predicted by the Dysthe theory. They also exceed the values obtained for near breaking waves in the laboratory experiment.

%%%%%%%%  Figure 4  %%%%%%%%%%
\begin{figure}[h]
\centering
\includegraphics[width=8.5cm]{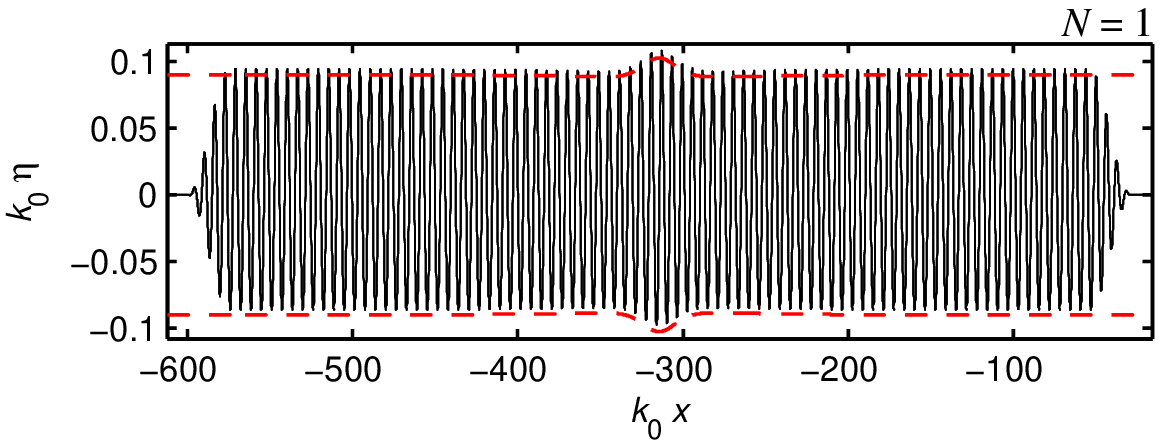}
\includegraphics[width=8.5cm]{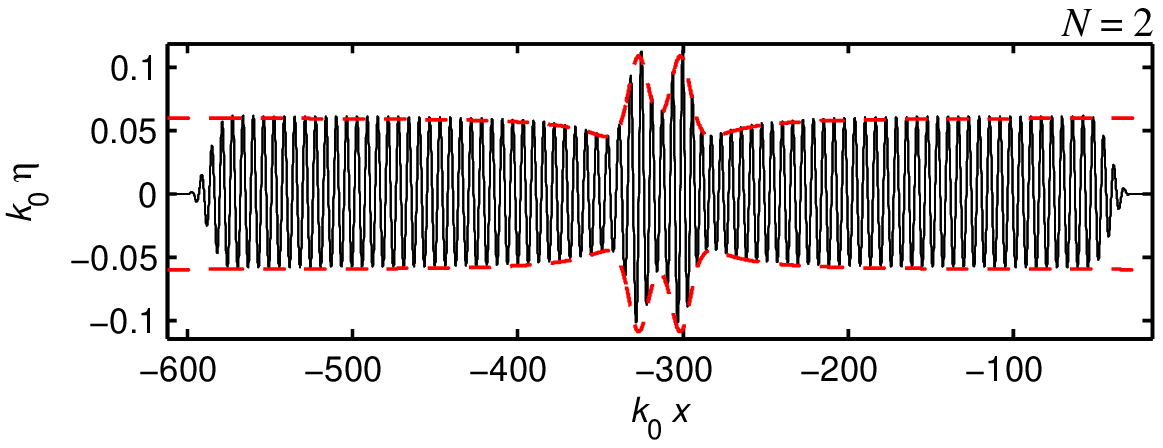}
\includegraphics[width=8.5cm]{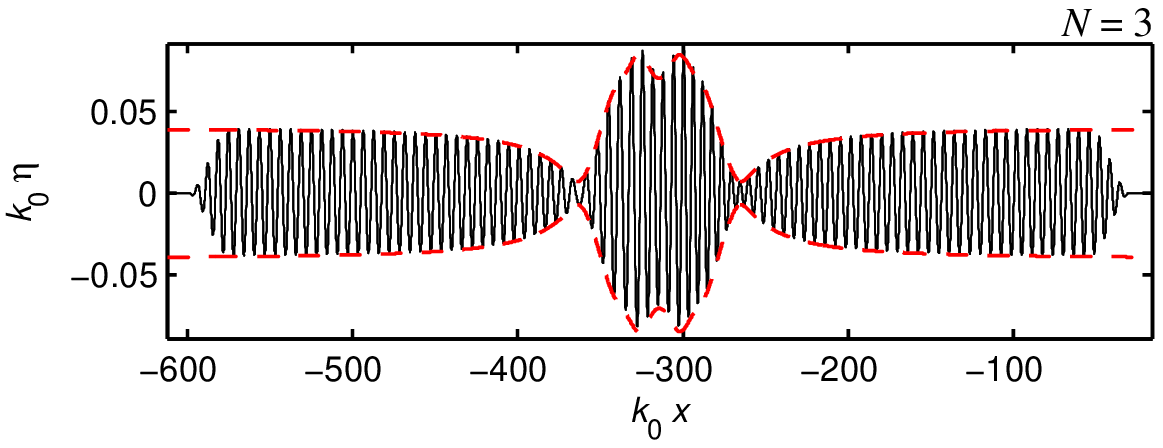}
\includegraphics[width=8.5cm]{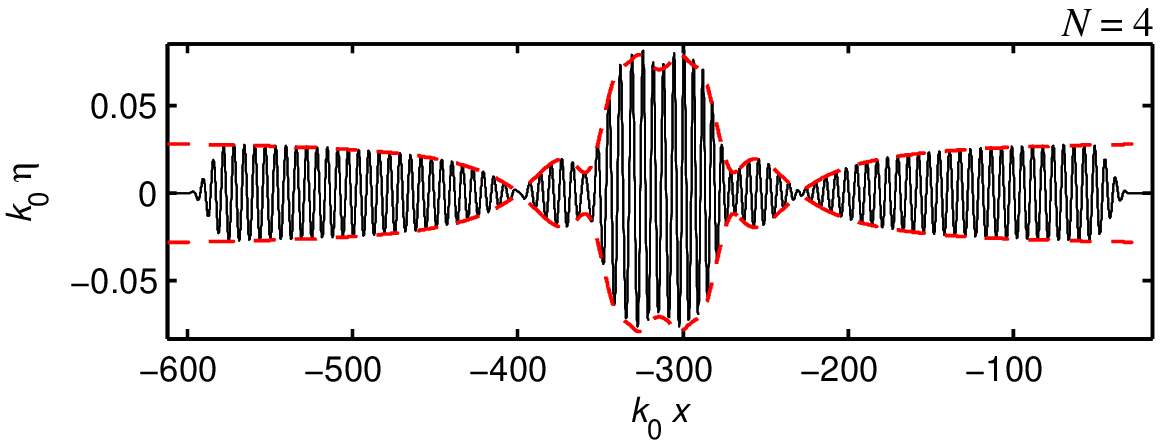}
\includegraphics[width=8.5cm]{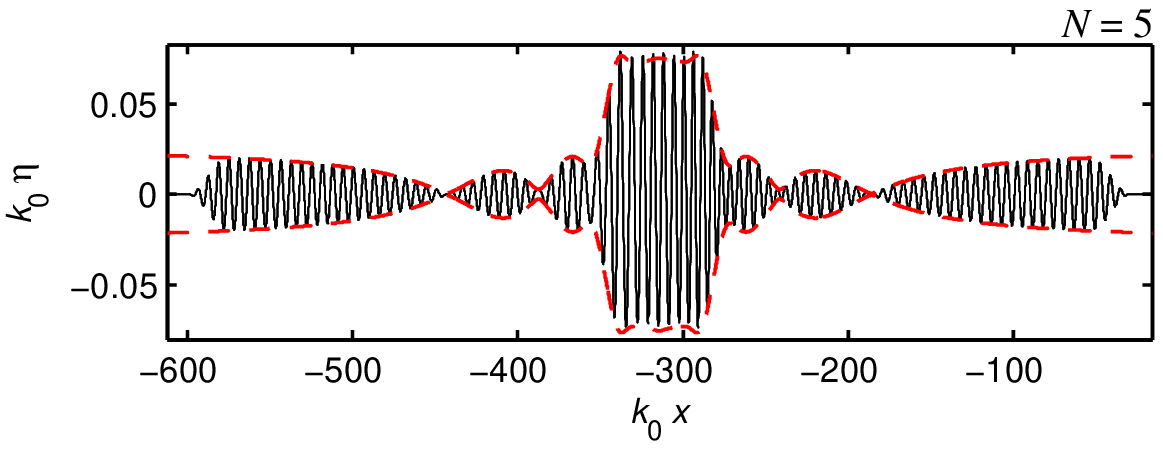}
\caption{(Color online) Initial conditions for the fully nonlinear simulations. Surface elevation is shown by the solid curves. The envelopes described by the corresponding analytic solutions of the NLS equation are shown by the dashed (red) curves. Each wavetrain is calculated 100 wave periods in advance of the wave focusing, $\tau=100~T_0$. The background wave steepness is highlighted in bold in corresponding lines of Table \ref{tab2}.}\label{fig4}
\end{figure}

The corrected maximum amplification of waves shown in Fig.3 is around 3-4 times independent of the order of the rational solution $N$. Higher wave amplifications observed in fully nonlinear simulations in comparison to the weakly nonlinear simulations can be explained by the strongly nonlinear nature of the focused waves.
The wavetrains at the time of the highest amplification are shown in Figs.5 and 6. These are also presented for all five values of $N$. The initial conditions for these simulations are shown in Fig.4. They are chosen at 100 wave periods prior the focusing point.  The initial wave steepness in all cases is close to the breaking point. These are shown in bold in Table \ref{tab2}.

%%%%%%%%  Figure 5  %%%%%%%%%%
\begin{figure}[h]
\centering
\includegraphics[width=8.5cm]{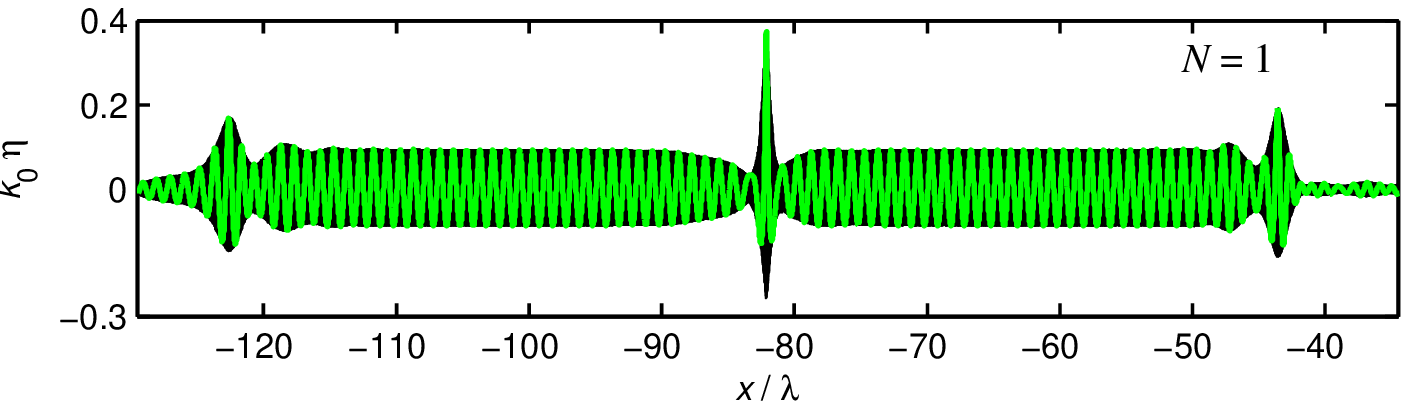}
\includegraphics[width=8.5cm]{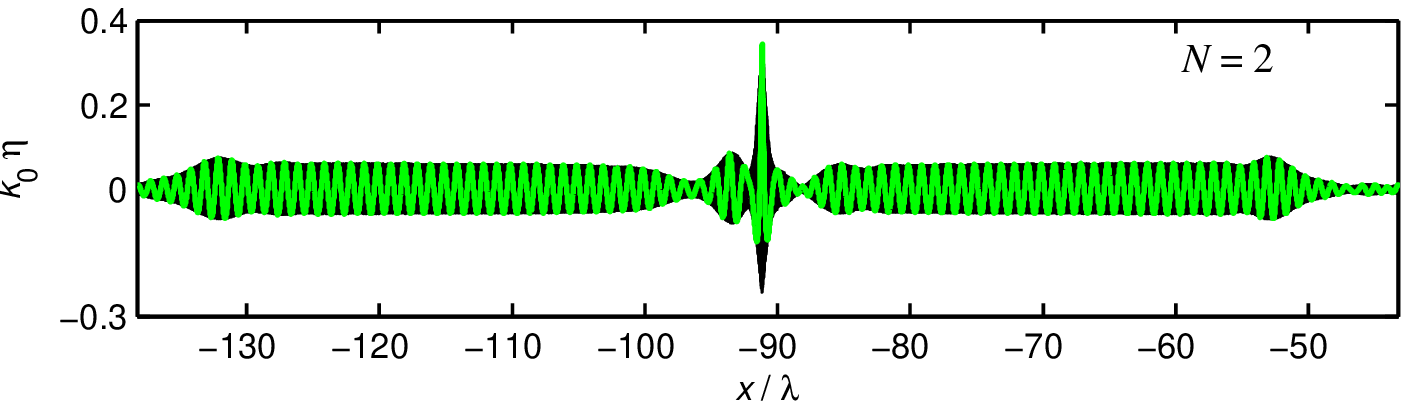}
\includegraphics[width=8.5cm]{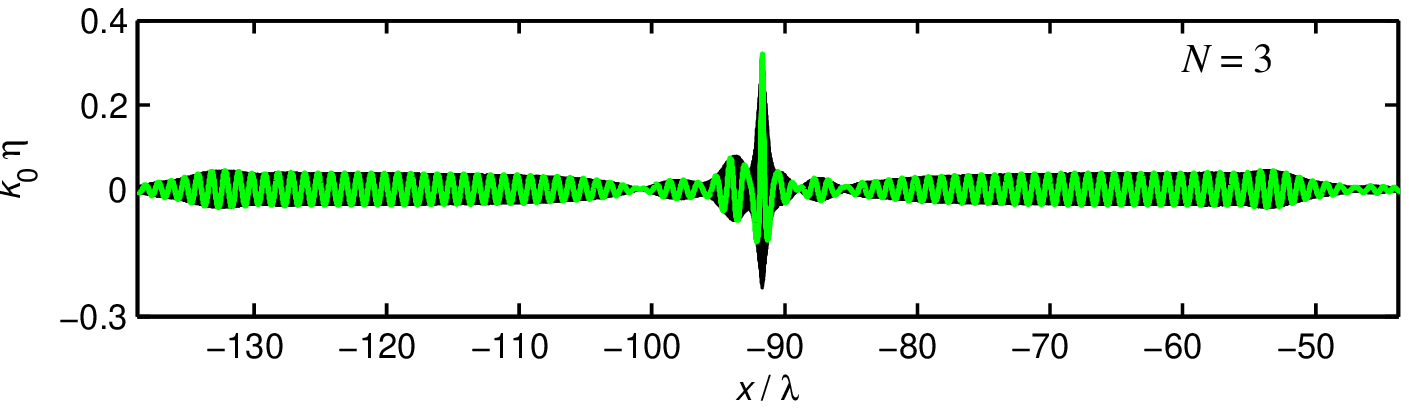}
\includegraphics[width=8.5cm]{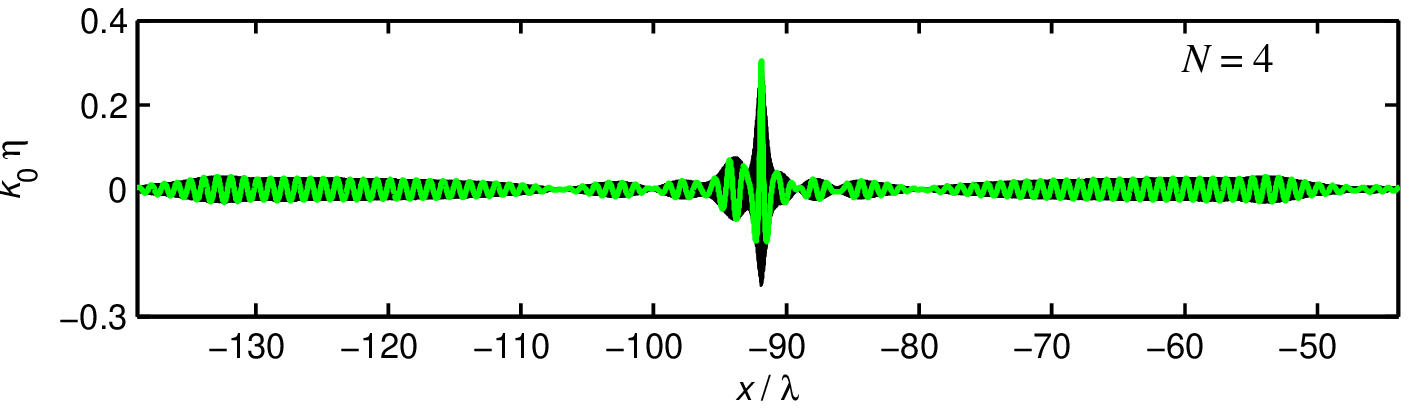}
\includegraphics[width=8.5cm]{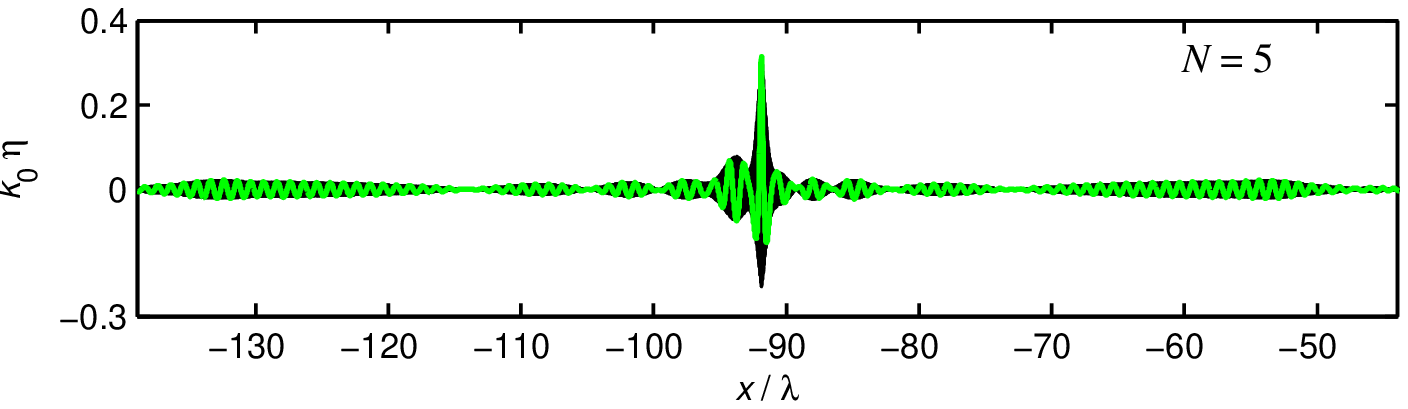}
\caption{(Color online) Wavetrains at the point of the maximum wave amplification obtained in the fully nonlinear simulations with initial conditions shown in Fig.4. The bright curves show the surface elevation while the grey area fills the envelope.}\label{fig5}
\end{figure}

The initial conditions shown by solid curves in Fig.4 correspond to the surface elevation in scaled coordinates. The dashed curves show the analytic NLS solutions for the wave envelope. The Stokes waves have vertical asymmetry making the wave crests slightly higher than the upper envelope, and the wave troughs shallower. This effect is less pronounced for waves with smaller amplitudes. For larger $N$, the initial condition have more complicated shape.

%%%%%%%%  Figure 6  %%%%%%%%%%
\begin{figure}[h]
\centering
\includegraphics[width=8.5cm]{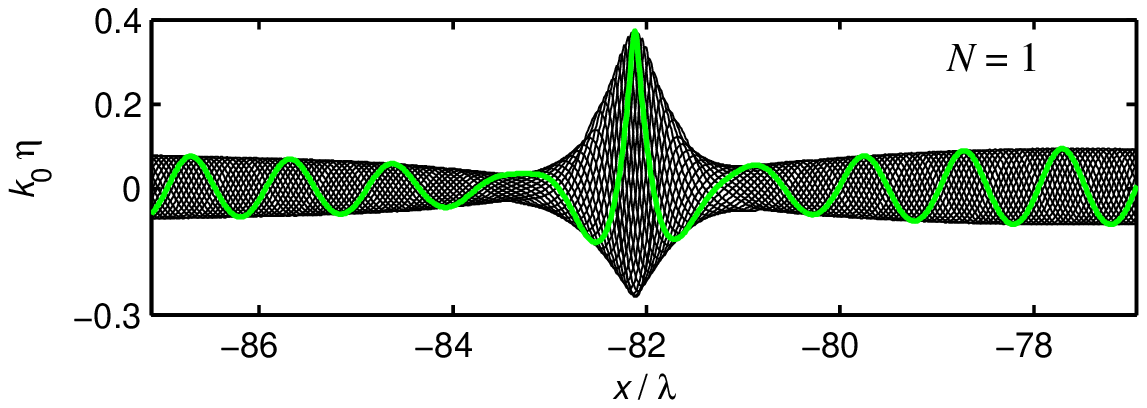}
\includegraphics[width=8.5cm]{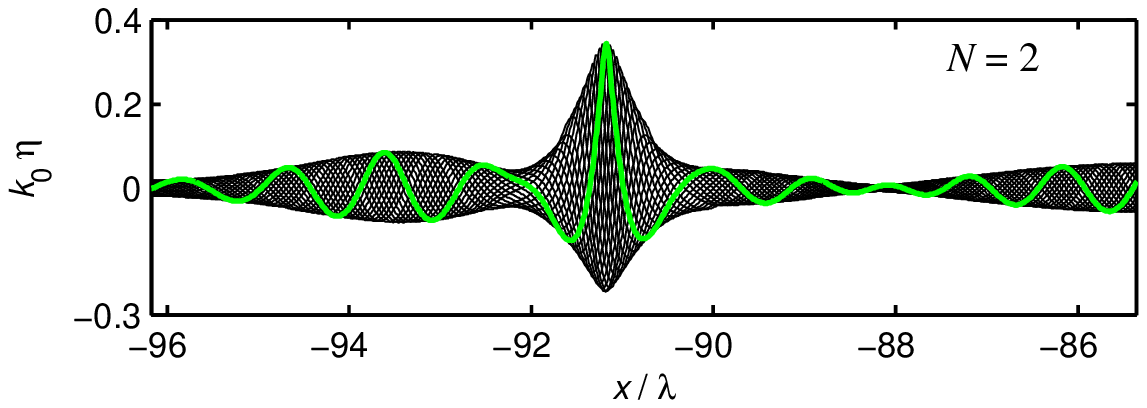}
\includegraphics[width=8.5cm]{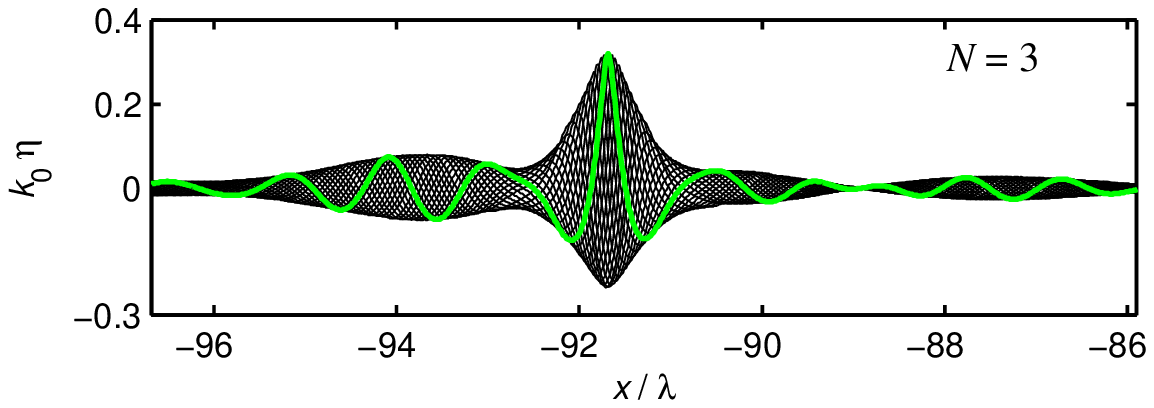}
\includegraphics[width=8.5cm]{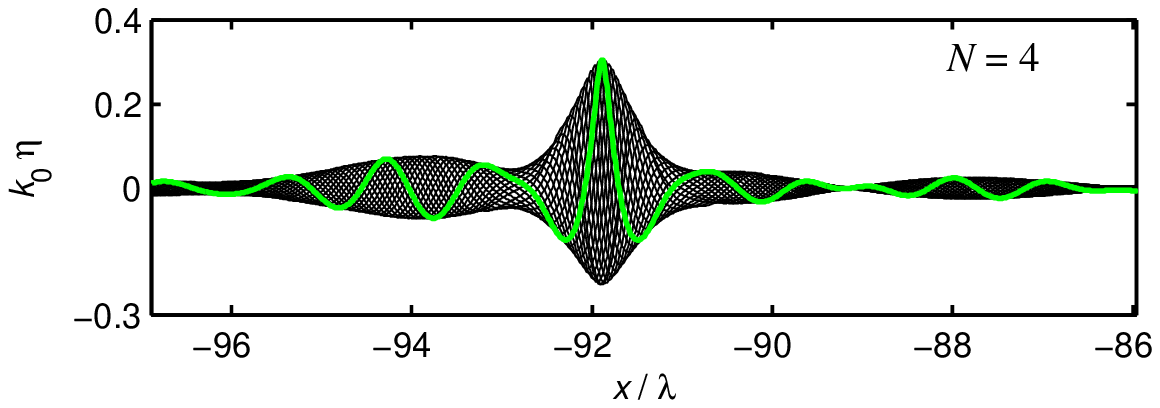}
\includegraphics[width=8.5cm]{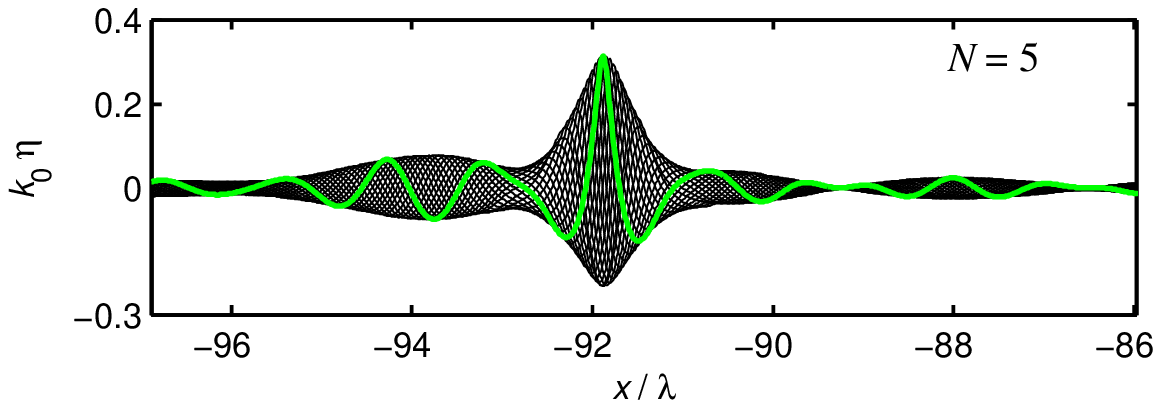}
\caption{(Color online) The central region of the curves shown in Fig. 5 in a magnified scale.}\label{fig6}
\end{figure}
	
The bright curves in Fig.5 show the surface elevations at the moment
when the maximum surface displacement is reached. The boundary of the hatched area clearly emphasises the envelope of the wave. The hatching itself is done with several surface elevation curves slightly shifted in phase. The central region of the curves is zoomed and shown separately in Fig.6. The main surface elevation curve here is contrasted in green while hatching is done by the black phase shifted lines just like in Fig.5.
Similar extreme wave trains were observed in simulations \cite{Slunyaev2012b}, and very much similar solitary wave groups were measured recently in laboratory conditions \cite{Slunyaev2012}.

To summarise, there is a set of wave packets with high amplitude central points that can be generated due to the nonlinear focusing. Just like in the NLS case, we can generate wave envelope structures with increasing number of nodes when increasing the values of $N$. Five lowest order wave profiles are shown in Fig.5. These can be predicted based on NLS results although there is no complete correspondence with the NLS solutions. As expected, the asymmetry is the main difference from the NLS case. Such asymmetry has been clearly observed in the experiments \cite{Chabchoub2012c}.
 In each case, the central region of the packet contains only a few wave oscillations. The central high amplitude part of the wave packet is another qualitative feature that in principle confirms the NLS prediction. Thus, the plots in Figs.5 and 6 generated in our numerical simulations qualitatively agree with the higher-order NLS rogue wave solutions as well as with their observations in the laboratory \cite{Chabchoub2012c}.

	An extensive and detailed study of modulated wave groups with high amplitude central peak is performed in \cite{Slunyaev2012b}. One of the main conclusions made in \cite{Slunyaev2012b} is that the nonlinearly focused waves, which are close to the wave breaking limit, have certain common features. The set of numerical simulations presented here also confirms this general conclusion. Firstly, the waves with maximum surface elevation shown by the thick green lines in Fig.6 have narrower crests than the Stokes wave. Secondly, the
crest tips have the shape of the steepest Stokes wave (see details in \cite{Slunyaev2012b}).

\section{Conclusions}

The goal of our present work is to study the hierarchy of rogue wave solutions given by the  exact solutions of the integrable NLS equation in real life. We concentrated on solutions of this set with lowest order from $N=1$ to $N=5$.
These solutions were recently observed experimentally in a water wave tank \cite{Chabchoub2012c}.
The observations were found to agree reasonably well with the analytic solutions despite the limitations
of the laboratory facility. This laboratory study motivated the research which is reported in the present paper.

Namely, here we try to avoid the experimental limitations related to the short length of the wave tank
and present simulations that describe the rogue waves with higher accuracy and in longer evolution than the experimental tank allows. Using two different approaches, we reproduced exact solutions of the integrable NLS equation. In one set of simulations we used Dysthe equations while the second technique is based on direct simulations
of the Euler equations.

Like in the experiment \cite{Chabchoub2012c}, the starting conditions in our simulations are specified according to the exact NLS solutions. For higher accuracy, we took into account a few bound wave components. Further evolution then is compared with the NLS predictions. One of the main questions addressed by the simulations is how the growth of the rogue wave until the maximum amplitude is described in the two approaches. The main advantage of the
strongly nonlinear simulations
is the ability to describe the evolution near the wave breaking limit.

In general, we confirm that the wave evolution in new numerical simulations follows the exact breather solutions of the NLS equation reasonably well. The following features have been observed.
\begin{enumerate}
\item There is a qualitative difference between the lowest $N=1$ and higher-order $N>1$ solutions. As expected, the dynamics described by the Peregrine breather ($N=1$) is much more robust than the dynamics prescribed by the solutions for $N>1$.

\item The distance required for focusing in the simulations is longer than predicted by the analytic solution in case $N=1$. For higher-order rogue waves $N>1$ the focusing distances in simulations and in theory are in good agreement.

\item For the lowest order rogue wave $N=1$ the envelope amplification of three is well confirmed by our numerical simulations. On the other hand, solutions with $N>1$ provide noticeably smaller amplification than prescribed by the NLS theory. Even the wave evolution may differ if the focusing distance is too long.

\item The effect of bound wave components is found to be very important for the process of focusing. Due to the contribution of bound waves, the wave crests are higher than the depth of wave troughs. Due to this difference,  the enhancement of wave amplitudes is effectively larger than without the bound waves. In experiments, the amplification is usually determined based on measurements of the wave crest amplitudes.

\item Longer focusing distances lead to distortions of nonlinear wave focusing. As a result, the actual wave maxima for $N>1$ does not noticeably exceed the value, which is observed in the case $N=1$.  The fully nonlinear simulations of nearly breaking waves show the maximum wave crest amplification up to about 4 times of the background. This estimate is in agreement with the results of numerical simulations in \cite{Tanaka1990} and \cite{Slunyaev2012b}. Consequently, the analytic NLS solutions with $N>1$ are not optimal with respect to the maximal wave enhancement for use as initial conditions for nearly breaking waves. For waves with smaller amplitudes, the agreement with the NLS theory may be better. On the other hand, then the growth time or distance increases as $\sim \varepsilon^{-2}$, where $\varepsilon$ is the wave steepness.

\end{enumerate}

\section{Acknowledgments}
The authors acknowledge the support of the Volkswagen Stiftung.
A.~Sl. and A.~Se. acknowledge support from RFBR grants 11-02-00483 and 12-05-33087.
The research by A.~S. was supported by the EC 7th Framework Programme (PIIFR-GA-2009-909389), and the research by A.~Se. was supported by grant MK-5222.2013.5.
E.~P. acknowledges support from RFBR grant 11-05-00216, Austrian Science Foundation (FWF) under project P24671, and
Federal Targeted Program "Research and educational personnel of
innovation Russia" for 2009–2013.
M.~O. was supported by ONR grant N000141010991 and by the European Union under the project EXTREME SEAS
(SCP8-GA-2009-234175).
N.~A. acknowledges partial support of the Australian Research
Council (Discovery Project No. DP110102068). N.~A. is a winner of the Alexander von Humboldt Award.


\begin{thebibliography}{99}
% 01
\bibitem[Chabchoub et al. (2012c)]{Chabchoub2012c} A. Chabchoub, N. Hoffmann, M. Onorato, A. Slunyaev, A. Sergeeva, E. Pelinovsky, and N. Akhmediev, Observation of a hierarchy of up to fifth-order rogue waves in a water tank. Phys. Rev. E {\bf86}, 056601 (2012).
% 02
\bibitem[Slunyaev et al. (2011)]{Slunyaev2011} A. Slunyaev, I. Didenkulova, E. Pelinovsky, Rogue Waters. Contemporary Physics {\bf52}, 571--590 (2011).
% 03
\bibitem[Kharif et al. (2009)]{Kharif2009} C. Kharif, E. Pelinovsky, A. Slunyaev \textit{Rogue Waves in the Ocean}, Springer-Verlag Berlin Heidelberg (2009).
% 04
\bibitem[Dysthe et al. (2008)]{Dysthe2008} K. B. Dysthe, H. E. Krogstad, and P. Muller, Oceanic rogue waves. Annu. Rev. Fluid. Mech. {\bf40}, 287--310 (2008).
% 05
\bibitem[Onorato et al. (2001)]{onorato01} M. Onorato, A.R. Osborne, M. Serio and S. Bertone,
Freak wave in random oceanic sea states,
Phys. Rev. Lett.  {\bf 86}, 5831--5834 (2001).
%6
\bibitem[Onorato et al. (2006)]{ONO06} M. Onorato, A. Osborne, M. Serio, L. Cavaleri, C. Brandini, C.T. Stansberg, Extreme waves, modulational instability and second order theory: wave flume experiments on irregular waves, Europ. J. Mech. B/Fluids, {\bf 25} 586--601 (2006)
% 06
\bibitem[Zakharov (1968)]{Zakharov1968}V. E. Zakharov, Stability of periodic waves of finite amplitude on a surface of deep fluid, J. Appl. Mech. Tech. Phys. {\bf2}, 190 (1968).
% 07
\bibitem[Zakharov \& Shabat (1972)]{Zakharov1972} V. E. Zakharov, V. E. and A. B. Shabat Exact theory of two-dimensional self-focussing and one-dimensional self-modulation of waves in nonlinear media. Sov. Phys. JETP {\bf34}, 62--69 (1972).
% 08
\bibitem[Osborne (2010)]{Osborne2010}A. Osborne, \textit{Nonlinear Ocean Waves and the Inverse Scattering Transform} (Academic Press, 2010).
% 09
\bibitem[Chabchoub et al. (2011)]{Chabchoub2011} A. Chabchoub, N. P. Hoffmann, and N. Akhmediev, Rogue wave observation in a water wave tank. Phys. Rev. Lett. {\bf106}, 204502 (2011).
% 09
\bibitem[Chabchoub et al. (2012a)]{Chabchoub2012a} A. Chabchoub, A., N. Hoffmann, M. Onorato, and N. Akhmediev, Super rogue waves: observation of a higher-order breather in water waves. Phys. Rev. X. {\bf2}, 011015 (2012).
% 10
\bibitem[Slunyaev et al. (2012)]{Slunyaev2012} A. Slunyaev, G. F. Clauss, M. Klein, M. Onorato, Simulations and experiments of 'Euler solitons' in surface water waves. Under revision for Phys. Fluids (2012)
% 11
\bibitem[Peregrine (1983)]{Peregrine1983} D. H. Peregrine, Water waves, nonlinear Schr\"{o}dinger equations and their solutions J. Austral. Math. Soc. Ser. B {\bf25}, 16--43 (1983).
% 12
\bibitem[Shrira \& Geogjaev (2010)]{Shrira2010} V. I. Shrira and V. V. Geogjaev, What makes the Peregrine soliton so special as a prototype of freak waves? J. Eng. Math. {\bf67}, 11--22 (2010).
% 13
\bibitem[Kuznetsov (1977)]{Kuznetsov1977} E. A. Kuznetsov, Solitons in a Parametrically Unstable Plasma, Sov. Phys. Dokl. {\bf22}, 507--508 (1977).
% 14
\bibitem[Akhmediev et. al (1987)]{Akhmediev1987} N. Akhmediev, V. M. Eleonskii, and N. E. Kulagin, Exact first-order solutions of the
nonlinear Schr\"{o}dinger equation. Theor. Math. Phys. USSR {\bf72}, 809--818 (1987).
% 15
\bibitem[Akhmediev et. al (1985)]{Akhmediev1985} Akhmediev, N., V. M. Eleonskii, N. E. Kulagin, Generation of periodic trains of
picosecond pulses in an optical fiber: exact solutions. Sov. Phys. JETP {\bf62}, 894--899 (1985).
% 16
\bibitem[Akhmediev et. al (2009a)]{Akhmediev2009a} N. Akhmediev, A. Ankiewicz, and M. Taki, Waves that appear from nowhere and disappear without a trace. Phys. Lett. A {\bf373}, 675--678 (2009).
% 17
\bibitem[Akhmediev et. al (2009b)]{Akhmediev2009b} N. Akhmediev, A. Ankiewicz, and J. M. Soto-Crespo, Rogue waves and rational solutions of the nonlinear Schrodinger equation. Phys. Rev. E {\bf80}, 026601 (2009).
% 18
\bibitem[Dubard et al. (2010)]{Dubard2010} P. Dubard, P. Gaillard, C. Klein and V. B. Matveev, On multi-rogue wave solutions of the NLS equation and positon solutions of the KdV equation,
in European Physical Journal, Special Topics, Discussion and Debate, Rogue Waves – Towards a Unifying Concept (Eds. N. Akhmediev and E. Pelinovsky) {\bf185}, pp. 247--258 (2010).
% 19
\bibitem[Gaillard (2012)]{Gaillard2012} P. Gaillard, Wronskian representation of solutions of the NLS equation and higher Peregrine breathers. Journal of Mathematical Sciences: Advances and Applications {\bf13}, 71--153 (2012).
% 20
\textcolor{red}{\bibitem[Trulsen (2006)]{Trulsen2006} K. Trulsen, Weakly nonlinear and stochastic properties of ocean wave fields: application to an extreme wave event. In: Waves in geophysical fluids: Tsunamis, Rogue waves, Internal waves and Internal tides. Eds.: Grue, J. \& Trulsen, K. CISM Courses and Lectures No. 489, (Springer, NY, Wein, 2006).}
% 21
\bibitem[Chabchoub et al. (2012b)]{Chabchoub2012b} Chabchoub, N. Akhmediev A., and N. P. Hoffmann, Experimental study of spatiotemporally localized surface gravity water waves, Phys. Rev. E. {\bf86}, 016311 (2012).
% 21
\bibitem[Trulsen \& Dysthe (1996)]{Trulsen1996} K. Trulsen, K. B. Dysthe, A modified nonlinear Schrödinger equation for broader bandwidth gravity waves on deep water, Wave Motion {\bf24}, 281 (1996).
% 22
\bibitem[Dysthe (1979)]{Dysthe1979} K. B. Dysthe, Note on a modification to the nonlinear Schrödinger equation for application to deep water waves. Proc. Roy. Soc. London A {\bf369}, 105--114 (1979).
% 23
\bibitem[Shemer et al. (2010)]{Shemer2010} L. Shemer, A. Sergeeva, A. Slunyaev, Applicability of envelope model equations for simulation of narrow-spectrum unidirectional random field evolution: experimental validation. Phys. Fluids {\bf22}, 016601 (2010).
% 24
\bibitem[Slunyaev (2005)]{Slunyaev2005} A.V. Slunyaev, A high-order nonlinear envelope equation for gravity waves in finite-depth water. JETP {\bf101}, 926--941 (2005).
% 25
\bibitem[West et al. (1987)]{West1987} B. J. West, K. A. Brueckner, R. S. Janda, D. M. Milder, and R.I. Milton, A new numerical method for surface hydrodynamics, J. Geophys. Res. {\bf92}, 11803 (1987).
% 26
\bibitem{Chalikov}
D. Chalikov and D. Sheinin, in Advances in Fluid Mechanics, edited by
W. Perrie, (Computational Mechanics, Southampton, 1998), Vol. 17, pp. 207--258.
% 27
\bibitem{Euler}
V. E. Zakharov, A. I. Dyachenko, and A. O. Prokofiev, New method for numerical simulation of a nonstationary potential flow of incompressible fluid with a free surface, Eur. J. Mech. B / Fluids, {\bf 21}, 283 (2002).
% 28
\bibitem[Slunyaev (2009)]{Slunyaev2009} A.V. Slunyaev, Numerical simulation of "limiting" envelope solitons of gravity waves on deep water. JETP {\bf109}, 676--686 (2009).
% 28
\bibitem[Slunyaev \& Shrira (2012)]{Slunyaev2012b} A. Slunyaev, V. Shrira The problem of the highest wave in a group: Fully nonlinear simulations of water wave breathers vs weakly nonlinear theory. Abstracts IUTAM Symp. "Waves in fluids: Effects of non-linearity, rotation, stratification and dissipation", 152--157 (2012).
A. Slunyaev and V. Shrira, On the highest non-breaking wave in a group: fully nonlinear water wave breathers vs weakly nonlinear theory. Submitted to J. Fluid Mech.
% 29
\bibitem[Tanaka (1990)]{Tanaka1990} M. Tanaka, Maximum amplitude of modulated wavetrain. Wave Motion {\bf12}, 559--568 (1990).

\end{thebibliography}
\end{document}